\documentclass[twocolumn,showpacs,preprintnumbers,amsmath,amssymb]{revtex4}

\usepackage{color}
\usepackage{graphicx}
\usepackage{dcolumn}
\usepackage{bm}

\begin{document}


\title{Long-time asymptotics 
and conservation laws in integrable systems}

\author{M.S. Hawkins}
\author{M.W. Long}
\affiliation{
School of Physics, Birmingham University, Edgbaston, Birmingham, B15 2TT, 
United Kingdom.
}
\author{X. Zotos}
\affiliation{
Department of Physics, University of Crete \\ and Foundation
for Research and Technology-Hellas, \\P. O. Box 2208, 71003 Heraklion,
Crete, Greece.}

\date{\today}

\begin{abstract}
One dimensional systems sometimes show pathologically slow decay of currents.  
This robustness can be traced to the fact that an {\it integrable} model is 
nearby in parameter space.  In integrable models some part of the current can 
be conserved, explaining this slow decay.  Unfortunately, although this 
conservation law is formally anticipated, in practice it has been difficult to 
find in concrete cases, such as the Heisenberg model.  We investigate this 
issue both analytically and numerically and find that the appropriate 
conservation law can be a {\it non-analytic} combination of the known local 
conservation laws and hence is invisible to elementary assumptions.
\end{abstract}

\maketitle

\section{\label{sec:level1}Introduction}

Although integrable systems are well understood in classical physics, their 
quantum counterparts are much more mysterious.  In classical physics we have 
phase-space, and an integrable system with 2$N$-dimensional phase-space can be 
characterised by a decomposition into a set of independent $N$-dimensional tori 
on which the trajectories are constrained, circling each `handle' independently 
with a characteristic frequency\cite{1}.  This reduction in the motion of the 
trajectories is accomplished by {\it conservation laws,} of which there are 
$N$, the values of which can be used to annotate the distinct tori.  In quantum 
physics no such generic picture exists, although some commonality between 
systems is emerging.  The majority of integrable systems are solved using the 
Bethe Ansatz\cite{ba} which has provided a collection of `non-trivial' systems 
which have been widely studied\cite{ba}.  
As well as the spectral properties, in 
a parallel investigation the conservation laws have been found:  Indeed, a 
countable number of conservation laws can be generated by a `boost or ladder 
operator' where the subsequent conservation law can be generated via 
commutation of the boost operator with the previous law\cite{gm}. 
Integrable systems have quite surprising properties.

Various (more or less obscure but quite interesting) anomalies have been 
discovered.  The additional conservation laws mean that nearby eigenstates 
behave independently and the local eigenvalue statistics is seen to be 
Poisson and not Wigner-Dyson\cite{rmt} as it is in normal systems. 
Indeed, this 
property can be used as a test for integrability!\cite{level}  If a current is 
started in an integrable system then it does not decay away completely, even 
if you wait forever and even at {\it finite} temperature!\cite{review}  
This final 
`anomaly' is even of experimental interest, because anomalously slow decay of 
currents has been observed in systems which have an integrable model nearby 
in parameter space.\cite{exp}

It is this final issue of long-time residuals that we examine here, asking the 
question of whether the long-time residual is {\it saturated} by the standard 
conservation laws\cite{gm} or whether there might be a secondary mechanism or 
perhaps even `non-local conservation laws'\cite{nonlocal} which contribute. 
Some 
aspects of these issues have previously been addressed but have led to the 
current confusion:  It has been shown that {\it formally} the long-time 
residuals are {\it necessarily} interpretable as being due to conservation 
laws\cite{suzuki} and yet for a particular system with a `known' non-vanishing 
residual\cite{z,kluemper} {\it all} the conservation laws yield a vanishing 
contribution to this residual due to symmetry\cite{znp}. This conundrum is our 
target.  We cast doubt on the previous analysis by showing that it is {\it not} 
enough to take a complete set of mutually commuting symmetries to describe the 
residual, but then find that this problem is not encountered for the concrete 
example that we study in depth.

Our investigation comes in two pieces.  Firstly we examine the problem {\it 
formally,} re-deriving the original saturation idea and extending it to the 
presence of `mutually commuting' conservation laws and to the issue of {\it 
compatibility} of those laws with the current.  Secondly we examine a 
particular case numerically, showing how our ideas work in practice and 
shedding light on the previously mentioned conundrum.  We find it possible to 
saturate the residual, but the conservation law required is non-analytic in 
the natural local conservation laws prevalent in the literature.

\section{Formal Analysis}

The physical idea behind a long-time residual is that at some time, say $t$=0, 
a current is started and then at some much later time the question of how much 
of the current is still flowing is addressed.  If a finite fraction of the 
current remains flowing for all time then there is a {\it long-time residual.}  
This issue may be investigated through a calculation of
\begin{equation}
\langle \hat j(\infty )\hat j(0)\rangle \equiv \lim _{T\mapsto \infty }
\frac{1}{T}\int _0^Tdt\langle \hat j(t)\hat j(0)\rangle ,
\end{equation}
where we employ the Heisenberg representation
\begin{equation}
\hat O(t)=e^{i\hat Ht}\hat Oe^{-i\hat Ht}
\end{equation}
and the angled brackets denote a thermal average
\begin{equation}
\langle \hat O\rangle \equiv \frac{tr \left[ e^{-\beta \hat H}\hat O\right] }
{tr \left[ e^{-\beta \hat H}\right] } .
\end{equation}
Note that there is a {\it linear response} aspect to this issue and that an 
infinitesimal current is actually introduced and then propagates.  The 
time-average is a technical trick to facilitate the calculation, allowing 
fluctuating terms to be eradicated:  Only terms with an essentially static 
phase can contribute, others {\it averaging} to zero.

Initially we will work with a finite system at the {\it formal} level in 
order to clarify the relationship between this quantity and conservation 
laws.  We start out by using an eigenbasis which 
diagonalises the Hamiltonian, denoted by $\mid n,m\rangle $ and satisfying
\begin{equation}
\hat H\mid n,m\rangle =\epsilon _n\mid n,m\rangle .
\end{equation}
The label $n$ denotes the distinct eigenvalues and the label $m$ is currently 
an arbitrary degeneracy label.  As we shall see, degeneracy is the crucial 
element to the problem and how we deal with this constitutes a formal 
resolution of the issue.  In this basis we find that
\begin{eqnarray}
\langle \hat O\rangle =\frac{1}{Z}\sum _ne^{-\beta \epsilon _n}\sum _m
\langle n,m\mid \hat O\mid n,m\rangle \nonumber &&\\\equiv \frac{1}{Z}
\sum _ne^{-\beta \epsilon _n}\sum _mO_{nn}^{mm} , && 
\end{eqnarray}
with the partition function
\begin{equation}
Z=\sum _ne^{-\beta \epsilon _n}\sum _m1
\equiv \sum _nM_ne^{-\beta \epsilon _n} ,
\end{equation}
where we have introduced the degeneracy of the energy $\epsilon _n$, as $M_n$, 
and the matrix element
\begin{equation}
O_{nn'}^{mm'}\equiv \langle n,m\mid \hat O\mid n',m'\rangle .
\end{equation}
The long-time residual may now be calculated using this basis and we find
\begin{widetext}
\begin{eqnarray}
\langle \hat j(\infty )\hat j(0)\rangle \equiv \lim _{T\mapsto \infty }
\frac{1}{T}\int _0^Tdt\frac{1}{Z}\sum _ne^{-\beta \epsilon _n}\sum _m
\langle n,m\mid e^{i\epsilon _nt}\hat j\sum _{n'}e^{-i\epsilon _{n'}t}
\sum _{m'}\mid n',m'\rangle \langle n',m'\mid \hat j\mid n,m\rangle &&\nonumber 
\\=\frac{1}{Z}\sum _{nn'}\sum _{mm'}e^{-\beta \epsilon _n}j_{nn'}^{mm'}j_{n'n}
^{m'm}\lim _{T\mapsto \infty }\frac{1}{T}\frac{\left[ e^{i(\epsilon _n-
\epsilon _{n'})T}-1\right] }{i(\epsilon _n-\epsilon _{n'})}=\frac{1}{Z}
\sum _ne^{-\beta \epsilon _n}\sum _{mm'}j_{nn}^{mm'}j_{nn}^{m'm} , &&
\end{eqnarray}
\end{widetext}
where we employed the idea that the eigenspectrum was {\it discrete} and we 
have not taken the thermodynamic limit.

This quantity is to be compared with the initial current
\begin{widetext}
\begin{equation}
\langle \hat j(0)\hat j(0)\rangle =\frac{1}{Z}\sum _{nn'}e^{-\beta \epsilon _n}
\sum _{mm'}\langle n,m\mid \hat j\mid n',m'\rangle \langle n',m'\mid \hat j\mid 
n,m\rangle \equiv \frac{1}{Z}\sum _{nn'}e^{-\beta \epsilon _n}
\sum _{mm'}j_{nn'}^{mm'}j_{n'n}^{m'm} ,
\end{equation}
\end{widetext}
and, if we assume that $\hat j$ is Hermitian, the matrix elements between 
states of differing energy are lost in the long-time limit.  To achieve a 
long-time residual the current operator must have a macroscopic component 
diagonal in energy.

The next important issue is that of conservation laws and at its most basic 
this constitutes operators which commute with the Hamiltonian.  Employing the 
eigenbasis to represent the general operator
\begin{equation}
\hat O=\sum _{nn'}\sum _{mm'}\mid n,m\rangle O_{nn'}^{mm'}\langle n',m'\mid , 
\end{equation}
and recognising the Hamiltonian as
\begin{equation}
\hat H=\sum _n\sum _m\mid n,m\rangle \epsilon _n\langle n,m\mid , 
\end{equation}
we find that
\begin{equation}
\left[ \hat H,\hat O\right] =\sum _{nn'}\sum _{mm'}\mid n,m\rangle \left[ 
\epsilon _n-\epsilon _{n'}\right] O_{nn'}^{mm'}\langle n',m'\mid ,
\end{equation}
and so any operator which involves a unique energy provides a conservation 
law.  Indeed, any operator provides a conserved piece
\begin{equation}
\hat O_c\equiv \sum _n\sum _{mm'}\mid n,m\rangle O_{nn}^{mm'}\langle n,m'\mid ,
\end{equation}
and we can decompose any operator into a conserved and non-conserved piece
\begin{equation}
\hat O\equiv \hat O_c+\hat O_{\bar c} .
\end{equation}
Employing this decomposition we can verify Mazur's 
inequality\cite{mazur} and the saturation theorem\cite{suzuki}
\begin{eqnarray}
\langle \hat A(\infty )\hat B(0)\rangle \equiv \langle \hat A_c(\infty )
\hat B_c(0)\rangle +\langle \hat A_c(\infty )\hat B_{\bar c}(0)\rangle 
&&\nonumber \\+\langle \hat A_{\bar c}(\infty )\hat B_c(0)\rangle 
+\langle \hat A_{\bar c}(\infty )\hat B_{\bar c}(0)\rangle . &&
\end{eqnarray}
Now since the conserved operators commute with the Hamiltonian, $\hat O_c(t)
=\hat O_c(0)\equiv \hat O_c$, we find that
\begin{equation}
\langle \hat A_c\hat B_c\rangle =\frac{1}{Z}\sum _ne^{-\beta \epsilon _n}
\sum _{mm'}A_{nn}^{mm'}B_{nn}^{m'm} ,
\end{equation}
and the first term reproduces the previous long-time residual expression when 
$\hat A=\hat j=\hat B$.  The second and third terms vanish because they are 
off-diagonal and the final term vanishes because it involves non-trivial time 
dependence.  We find that
\begin{equation}
\langle \hat j(\infty )\hat j(0)\rangle =\frac{\langle \hat j(0)\hat j_c\rangle 
\langle \hat j_c\hat j(0)\rangle }{\langle \hat j_c\hat j_c\rangle } ,
\end{equation}
and further, for a set of orthogonal conservation laws $\hat C_a$ which 
satisfy
\begin{equation}
\langle \hat C_a^\dagger \hat C_{a'}\rangle =\delta _{aa'}\langle \hat C_a
^\dagger \hat C_a\rangle ,
\end{equation}
then
\begin{equation}
\langle \hat j(\infty )\hat j(0)\rangle \ge\sum _a\frac{\langle \hat j(0)\hat 
C_a\rangle \langle \hat C_a^\dagger \hat j(0)\rangle }{\langle \hat C_a
^\dagger \hat C_a\rangle } ,
\label{mazur}
\end{equation}
which is Mazur's inequality. This can be proved by representing
\begin{equation}
\hat j_c\equiv \sum _a\alpha _a\hat C_a+\Delta \hat j_c ,
\end{equation}
with
\begin{equation}
\langle \hat C_a^\dagger \Delta \hat j_c\rangle =0 \; \; \; \; \; \Rightarrow 
\; \; \; \; \; \alpha _a=\frac{\langle \hat C_a^\dagger \hat j_c\rangle }
{\langle \hat C_a^\dagger \hat C_a\rangle }=\frac{\langle \hat C_a^\dagger \hat 
j(0)\rangle }{\langle \hat C_a^\dagger \hat C_a\rangle } ,
\end{equation}
and then
\begin{widetext}
\begin{equation}
\langle \hat j(\infty )\hat j(0)\rangle -\sum _a\frac{\langle \hat j(0)\hat C_a
\rangle \langle \hat C_a^\dagger \hat j(0)\rangle }{\langle \hat C_a
^\dagger \hat C_a\rangle }=\langle \hat j_c\hat j_c\rangle -\sum _a\alpha 
_a^\dagger \langle \hat C_a^\dagger \hat C_a\rangle \alpha _a
=\langle \left( \hat j_c-\sum _a\alpha _a^\dagger \hat C_a^\dagger \right) \! 
\left( \hat j_c-\sum _a\alpha _a\hat C_a\right) \rangle =\langle \Delta \hat 
j_c^\dagger \Delta \hat j_c\rangle \ge 0 .
\end{equation}
\end{widetext}
The identity is the dominant content of the argument by Suzuki\cite{suzuki} 
and indeed 
shows that long-time residues may be interpreted as stemming from 
conservation laws in this broad sense.

In integrable systems there are a collection of non-trivial conservation laws 
that are thought to play a similar role to those in classical systems.  The 
current issue is whether Mazur's inequality\cite{mazur} can be {\it exhausted} 
by these laws and the conserved quantity, $\hat j_c$, can be written down as a 
{\it function} of these conservation laws.

The initial issue is as to which conserved operators are accessible to these 
conservation laws.  We expect to be dealing with mutually commuting operators 
which include the Hamiltonian
\begin{equation}
\left[ \hat C_a,\hat C_{a'}\right] =0 ,
\end{equation}
with $\hat C_0\equiv \hat H$.  Since any collection of commuting Hermitian 
operators may be simultaneously diagonalised, we are led to a restricted 
basis, $\mid n,c,m\rangle $, which satisfies
\begin{equation}
\hat C_a\mid n,c,m\rangle =\epsilon _{nc}^a\mid n,c,m\rangle , \; \; \; \; \; 
\; \; \; \epsilon _{nc}^0\equiv \epsilon _n ,
\end{equation}
where for each choice of the distinct pair $\{ n,c\} $ and $\{ n',c'\} $ we 
have an $a$ for which $\epsilon _{nc}^a\neq \epsilon _{n'c'}^a$ and the states 
are non-degenerate in this generalised sense.  Now clearly
\begin{equation}
\hat C_a\equiv \sum _{ncm}\mid n,c,m\rangle \epsilon _{nc}^a\langle n,c,m\mid 
\equiv \sum _{nc}\epsilon _{nc}^a\hat P_{nc} ,
\end{equation}
and each operator can be constructed from the projection operators
\begin{equation}
\hat P_{nc}\equiv \sum _m\mid n,c,m\rangle \langle n,c,m\mid .
\end{equation}
Further, it should be clear that
\begin{equation}
\hat P_{nc}\equiv \prod _{a\in A}\prod _{\epsilon ^a_{n'c'}\neq \epsilon ^a
_{n,c}}\left[ \frac{\hat C^a-\epsilon _{n'c'}^a}{\epsilon _{nc}^a-\epsilon 
_{n'c'}^a}\right] 
\end{equation}
for all of the operators described by set $A$.  So, each projection operator 
can be created from the conservation laws and each conservation law can be 
created from the projection operators and the two operator subspaces are 
equivalent.  The most general conservation law generated by this class of 
conservation laws takes the form
\begin{equation}
\hat A_c\equiv \sum _{nc}\sum _m\mid n,c,m\rangle A_{nc}\langle n,c,m\mid 
\equiv \sum _{nc}A_{nc}\hat P_{nc} .
\label{project}
\end{equation}

Although the long-time residual may be generated from the conservation law 
$\hat j_c$, any particular set of conservation laws may only generate part 
of the residual.  For the subspace annotated by $nc$, we can only find the 
piece of $\hat j_c$ which is proportional to the identity operator and 
consequently the maximal contribution to the Mazur inequality (Eq.\ref{mazur}) 
from a set of conservation laws comes from the conservation law
\begin{equation}
\hat j_M\equiv \sum _{ncm}\mid n,c,m\rangle \frac{1}{M_{nc}}\sum _{m'}
j_{nc\; nc}^{m'\; \; m'}\langle n,c,m\mid  ,
\end{equation}
where $M_{nc}$ is the degeneracy of the appropriate subspace and the 
maximal Mazur contribution is
\begin{equation}
\langle \hat j(\infty )\hat j(0)\rangle _M\equiv \frac{1}{Z}\sum _{nc}e^{-\beta 
\epsilon _n}\frac{1}{M_{nc}}\sum _{mm'}j_{nc\; nc}^{m\; \; m}
j_{nc\; nc}^{m'\; \; m'} .
\end{equation}
If we only employ the Hamiltonian itself as the system of conservation laws, 
then we also find
\begin{equation}
\langle \hat j(\infty )\hat j(0)\rangle _H\equiv \frac{1}{Z}\sum _ne^{-\beta 
\epsilon _n}\frac{1}{M_n}\sum _{cc'}\sum _{mm'}j_{nc\; nc}^{m\; \; m}
j_{nc'\; nc'}^{m'\; \; m'} ,
\end{equation}
whereas the full long-time residual in this basis is
\begin{equation}
\langle \hat j(\infty )\hat j(0)\rangle \equiv \frac{1}{Z}\sum _ne^{-\beta 
\epsilon _n}\sum _{cmm'}j_{nc\; nc'}^{m\; \; m'}
j_{nc'\; nc}^{m'\; \; m} .
\end{equation}

We find a hierarchy of {\it four} quantities:
\begin{eqnarray}
\lefteqn{I_0=\langle \hat j(0)\hat j(0)\rangle }\nonumber \\&&=\frac{1}{Z}
\sum _ne^{-\beta \epsilon _n}\sum _{cm}\sum _{n'c'm'}j_{nc\; n'c'}^{m\; \; m'}
j_{n'c'\; nc}^{m'\; \; m} ,\\\lefteqn{I_1=\langle \hat j(\infty )\hat j(0)\rangle 
}\nonumber \\&&=\frac{1}{Z}\sum _ne^{-\beta \epsilon _n}\sum _{cm}\sum _{c'm'}
j_{nc\; nc'}^{m\; \; m'}j_{nc'\; nc}^{m'\; \; m} ,
\\\lefteqn{I_2=\langle \hat j(\infty )\hat j(0)\rangle _M}\nonumber 
\\&&=\frac{1}{Z}\sum _ne^{-\beta \epsilon _n}\sum _{cm}\frac{1}{M_{nc}}
\sum _{m'}j_{nc\; nc}^{m\; \; m}j_{nc\; nc}^{m'\; \; m'} ,
\\\lefteqn{I_3=\langle \hat j(\infty )\hat j(0)\rangle _H}\nonumber 
\\&&=\frac{1}{Z}\sum _ne^{-\beta \epsilon _n}\sum _{cm}\frac{1}{M_n}
\sum _{c'm'}j_{nc\; nc}^{m\; \; m}j_{nc'\; nc'}^{m'\; \; m'} ,
\end{eqnarray}
with $I_0\ge I_1\ge I_2\ge I_3$.  Note that for a Hermitian matrix, the 
identity
\begin{widetext}
\begin{equation}
\sum _{aa'}\left( H_{a'a}-\delta _{aa'}\frac{1}{A}\sum _bH_{bb}\right) \left( 
H_{aa'}-\delta _{aa'}\frac{1}{A}\sum _bH_{bb}\right) =\sum _{aa'}H_{a'a}H_{aa'}-
\frac{1}{A}\sum _{aa'}H_{aa}H_{a'a'}\ge 0
\end{equation}
\end{widetext}
completes the proof of the inequalities.

$I_0$ is the original current and the components of current which connect 
states with different energies are lost over time until we reach the 
long-time residual, $I_1$.  The contributions that connect different states 
at the same energy are lost when we reduce down to the Mazur contribution 
directly described by the conservation laws, $I_2$, but all the contributions 
from different $c$'s still contribute.  We find only the average contribution 
over the degeneracy label.  The final reduction to the Hamiltonian-only 
Mazur contribution involves an average over all the matrix elements at each 
energy, $I_3$.

We can now answer the question of when the {\it commuting} algebra of 
conservation laws is enough to describe the long-time residual of the 
current:  Formally, the long-time residual is exhausted by the conservation 
laws only when the current operator restricted to each energy subspace is 
{\it the identity} in the basis which diagonalises the symmetries.  Given that 
there is a conservation law which exhausts the long-time residual, $\hat j_c$, 
we still have the issue of when we manage to capture it in an arbitrary 
collection of conservation laws.  In the original argument\cite{suzuki} 
a ``maximal 
set of mutually commuting conservation laws" is used as a starting point.  
{\it This is not sufficient!} 
In general the `mutually commuting' is a technical 
problem:  Note that saturation fails when the relevant conservation laws are 
{\it non-Abelian,} such as
\begin{equation}
\mid 1\rangle \langle 1\mid \; \; \; \; \; \; \; \; \mid 2\rangle \langle 
2\mid \; \; \; \; \; \; \; \; \mid 1\rangle \langle 2\mid \; \; \; \; \; \; 
\; \; \mid 2\rangle \langle 1\mid 
\end{equation}
where only the first two are maximally mutually commuting, but the current 
operator could include some of the $\mid 1\rangle \langle 2\mid +\mid 2\rangle 
\langle 1\mid $ conservation law, which is inaccessible.

The degeneracy is clearly a crucial issue, since if the eigenbasis is 
non-degenerate then the labels $c$ and $m$ are irrelevant and we find that
\begin{equation}
I_1=I_3=\frac{1}{Z}\sum _ne^{-\beta \epsilon _n}j_{nn}j_{nn} ,
\end{equation}
and the Hamiltonian itself generates the long-time residual and the conserved 
piece of the current is an elementary function of the Hamiltonian
\begin{equation}
\hat j_c=\sum _n\mid n\rangle j_{nn}\langle n\mid \equiv 
\sum _nj_{nn}\hat P_n[\hat H] ,
\end{equation}
with
\begin{equation}
\hat P_n[\hat H]=\prod _{m\neq n}\frac{\hat H-\epsilon _m}
{\epsilon _n-\epsilon _m}
\end{equation}
the relevant projection operator.  Indeed, for this case the Hamiltonian 
generates the {\it only} conservation laws and the system should not be 
thought of as being integrable in any way similar to the classical analogues 
at all.

For any mutually commuting set of conservation laws we can construct a single 
conservation law from which all others may be constructed.  We may choose
\begin{equation}
\hat A_c\equiv \sum _{nc}\sum _m\mid n,c,m\rangle A_{nc}\langle n,c,m\mid 
\equiv \sum _{nc}A_{nc}\hat P_{nc} ,
\end{equation}
as in equation (\ref{project}), and if all the $A_{nc}$ are distinct then
\begin{equation}
\hat P_{nc}=\prod _{(n'c')\neq (nc)}\frac{(\hat A_c-A_{n'c'})}
{(A_{nc}-A_{n'c'})} ,
\end{equation}
and consequently
\begin{equation}
\hat C_a=\sum _{nc}\epsilon ^a_{nc}\hat P_{nc}\equiv \hat C_a[\hat A_c]
\end{equation}
is an elementary function of $\hat A_c$.  In a calculation of a Mazur 
inequality, in principle, we can replace a set of chosen conservation 
laws with powers of a {\it single law} which is an arbitrary linear 
superposition of the original, because such a law is generically 
non-degenerate with respect to the chosen set.

In classical problems we think in terms of phase-space and trajectories:  Each 
conservation law corresponds to a restriction on the dimensionality of the 
subspace that a trajectory is moving in, with each law corresponding to the 
loss of a dimension and a complete set of $N$ conservation laws in 
2$N$-dimensional phase-space leading to an $N$-torus remaining with the 
trajectories cycling endlessly about the $N$ handles of the torus at 
independent rates, for an integrable system.  In quantum problems the analogue 
is that an eigenvalue of an operator is fixed and the system is restricted 
to the projected subspace for which the eigenvalue is fixed.  The {\it 
degeneracy} in that subspace amounts to the residual freedom that remains to 
the system.  Any additional conservation laws which commute with the 
Hamiltonian can be simultaneously diagonalised leading to a reduction in the 
projected subspace that the system can be restricted to.  This analogy very 
much follows our development and we have analysed our long-time residuals in 
terms of these restrictions.  The natural quantum analogue to a classical 
integrable system is then one for which the simultaneous diagonalisation of 
{\it all} the conservation laws leads to a unique basis and no residual 
freedom.

For an integrable system we anticipate that the labels $n$ and $c$ are 
sufficient and that the label $m$ becomes redundant.  It is still not clear 
whether the conservation laws exhaust the long-time residual, because the 
current operator could connect states at the same energy but with different 
values of the other conservation laws.  This eventuality would correspond 
directly to the previous comment that even a complete set of mutually commuting 
laws does {\it not} guarantee to exhaust all non-Abelian conservation laws.  
We now move on to a particular case, and analyse these formal ideas for a 
concrete example in an attempt to resolve the failure of conservation laws to 
predict the long-time residual of the Heisenberg model\cite{z,kluemper}.

\section{Long-time residuals and conservation laws in the XXZ model}

The particular system that we use for the current numerical investigation is 
the XXZ model
\begin{equation}
\hat H=\sum _i\hat h_{i,i+1}=\sum _i\left( \hat S_i^x\hat S_{i+1}^x+\hat 
S_i^y\hat S_{i+1}^y+\Delta \hat S_i^z\hat S_{i+1}^z\right) ,
\end{equation}
which is integrable\cite{bethe}.  
For this model the conservation laws have been 
established\cite{gm}.  The appropriate boost operator is
\begin{equation}
\hat B=\sum _ii\hat h_{i,i+1} ,
\end{equation}
and a sequence of conservation laws can be generated by
\begin{equation}
\hat C_{n+1}=\left[ \hat B,\hat C_n\right] ,
\end{equation}
with $\hat C_0\equiv \hat H$, in notation consistent with our formal analysis.  
These conservation laws are {\it local,} in the sense that they are a sum of 
translated copies of interactions which involve a finite contiguous set of 
sites.  Indeed, each subsequent law involves an extension of the length of 
the interacting region by a single site.  For all the undeniable mathematical 
beauty of this construction, unfortunately, it is of no practical use to us!  
To investigate saturation, we have been forced into analysing {\it finite} 
systems where the number of conservation laws is controllable and the 
extensive nature of the boost operator makes it incompatible with the periodic 
boundary conditions necessary to our calculations.

We have employed a more cumbersome approach to creating the conservation laws.  
The construction involves the `R-matrix'
\begin{equation}
\hat R_{ij}(\lambda )=\frac{S(\lambda )}{2}+2\left( \hat S_i^x\hat S_j^x
+\hat S_i^y\hat S_j^y+C(\lambda )\hat S_i^z\hat S_j^z\right) ,
\end{equation}
where
\begin{widetext}
\begin{eqnarray}
\Delta =\cos \delta \le 1\; \; \; \; \; \; \; \; \; \; \; \; 
S(\lambda )=\frac{\sin \frac{\delta }{2}(1+\lambda )}{\sin \frac{\delta }{2}}
\; \; \; \; \; \; \; \; C(\lambda )=\frac{\cos \frac{\delta }{2}(1+\lambda )}
{\cos \frac{\delta }{2}}&&\nonumber \\ \Delta =1\; \; \; \; \; \; \; \; \; \; 
\; \; \; \; \; \; \; \; \; \; \; \; \; \; S(\lambda )=1+\lambda \; \; \; \; \; 
\; \; \; \; \; \; \; \; \; \; \; \; \; \; \; C(\lambda )=1\; \; \; \; \; \; 
\; \; \; \; \; \; &&\\\Delta =\cosh \delta \ge 1\; \; \; \; \; \; \; \; \; \; 
S(\lambda )=\frac{\sinh \frac{\delta }{2}(1+\lambda )}{\sinh \frac{\delta }{2}}
\; \; \; \; \; \; C(\lambda )=\frac{\cosh \frac{\delta }{2}(1+\lambda )}{\cosh 
\frac{\delta }{2}}&&\nonumber 
\end{eqnarray}
\end{widetext}
in terms of which one can construct the monodromy matrix\cite{ba} for $N$ sites 
with periodic boundary conditions
\begin{equation}
\hat T[\lambda ]=tr_0\left( \hat R_{01}[\lambda ]\hat R_{02}[\lambda ]...
\hat R_{0N}[\lambda ]\right) .
\end{equation}
The trace is over the spin ${\bf \hat S}_0$ which is a dummy spin.  
This matrix controls the conservation laws using the result
\begin{equation}
\left[ \hat T[\lambda ],\hat T[\lambda ']\right] =0 ,
\end{equation}
for all $\lambda $, $\lambda '\in ${\cal R}, which guarantees the linear 
independence of any operators that can be constructed from this matrix.  The 
{\it local} conservation laws are derived from
\begin{equation}
\ln \left( \hat T[\lambda ]\hat T[0]^{-1}\right) =\sum _{n=1}^\infty 
\frac{\lambda ^n}{n!}\hat C_{n-1} ,
\end{equation}
with $\hat C_0\equiv \hat H$, and all subsequent operators being independent.

The previous formal ideas are straightforward to analyse on a finite system:  
Employing a `dummy' spin to describe ${\bf \hat S}_0$ it is elementary to 
calculate the monodromy matrix as a function of both $\lambda $ and $\delta $.  
This real but non-Hermitian matrix may be diagonalised and is almost invariably 
(respecting $\hat S^z=\sum _i\hat S_i^z$) non-degenerate.  This provides an 
essentially unique basis in which all the conservation laws are diagonal.  The 
final degeneracy label in our analysis becomes irrelevant and our analysis 
reduces to
\begin{eqnarray}
I_0=\frac{1}{Z}\sum _ne^{-\beta \epsilon _n}\sum _{n'cc'}
j_{nn'}^{cc'}j_{n'n}^{c'c} , &&\\
I_1=\frac{1}{Z}\sum _ne^{-\beta \epsilon _n}\sum _{cc'}
j_{nn}^{cc'}j_{nn}^{c'c} , &&\\
I_2=\frac{1}{Z}\sum _ne^{-\beta \epsilon _n}\sum _c
j_{nn}^{cc}j_{nn}^{cc} , &&\\
I_3=\frac{1}{Z}\sum _ne^{-\beta \epsilon _n}\frac{1}{M_n}\sum _c
j_{nn}^{cc}\sum _{c'}j_{nn}^{c'c'} , &&
\end{eqnarray}

Our formal considerations involved an arbitrary operator, $\hat j$, which we 
designated a `current'.  For the XXZ-model there is a {\it physical} aspect to 
the current, which can be constructed from either an infinitesimal `twist' 
in the boundary conditions (corresponding to a magnetic field through the 
loop) or the local continuity equation
\begin{eqnarray}
\frac{\partial \hat S^z}{\partial t}({\bf x})+\nabla .{\bf j}^z({\bf x})=0
\; \; \; \; \; \; \; \Rightarrow &&\nonumber \\
\frac{1}{i\hbar}\left[ \hat H,\hat S_j^z\right] +\frac{1}{\Delta 
x}\left( \hat j^z_{j+\frac{1}{2}}-\hat j^z_{j-\frac{1}{2}}\right) =0 , &&
\end{eqnarray}
which is solved by
\begin{equation}
\hat j^z_{j+\frac{1}{2}}\propto \hat S^y_j\hat S^x_{j+1}-\hat S^y_j\hat S^x
_{j+1}.
\end{equation}
Up to a constant, this provides our chosen current operator
\begin{eqnarray}
\hat j=2\sum _j\left( \hat S_j^y\hat S_{j+1}^x-\hat S_j^x\hat S_{j+1}^y
\right) &&\nonumber \\=(-i)\sum _j\left( \hat S^+_j\hat S_{j+1}^-
-\hat S_j^-\hat S_{j+1}^+\right) . &&
\end{eqnarray}
We note that the Hamiltonian contribution vanishes for this 
operator.  This can be proved using the symmetry
\begin{equation}
\hat U\equiv \exp \left( i\pi \sum _j\hat S_j^x\right) ,
\end{equation}
a rotation through $\pi $ about the $x$-axis, which preserves the Hamiltonian 
but reverses the sign of $\hat j$.  In the basis that simultaneously 
diagonalises $\hat H$ and $\hat U$ we find that $\hat j$ is necessarily 
off-diagonal and consequently $I_3$ vanishes.  Note that $\hat U$ does not 
respect $\hat S^z$ but does respect the symmetries $\hat C^a$ and so does not 
impinge on the standard conservation laws.  Although in principle $I_1$ and 
$I_2$ are distinct, for this model we find that the two quantities agree 
exactly and in the unique natural basis the relevant matrix elements of the 
current are diagonal.

As yet we have not encountered our conundrum:  If we employ {\it all} the
conservation laws then in our formal sense we achieve {\it all} the long-time
residual for the current model.  It is not straightforward, in practice, to
employ the eigenbasis and one would like to employ a {\it single} conservation
law that has a non-trivial overlap with the current to demonstrate whether the
long-time residual exists at all, let alone is being exhausted.  It is this 
issue that has proven problematic for the current model, where at 
`half-filling' (corresponding to the subspace $\hat S^z$=0) it has been shown 
that {\it none} of the conservation laws can contribute to the long-time 
residual, essentially because they respect $\hat U$ for this filling and then 
any expectation value involving both the current and any combination of the 
local conservation laws vanishes, eliminating the contribution to the Mazur
inequality.

So far we have discussed the pure XXZ model, but there is a natural extension 
to include a magnetic field
\begin{equation}
H=H_0-B\sum _i\hat S^z_i ,
\end{equation}
which activates the local conservation laws and allows a non-trivial 
contribution to the long-time residual to be obtained from the simplest 
law\cite{znp}. 
This additional term commutes with the original Hamiltonian {\it 
and} all the local conservation laws and completes the uniqueness of the 
diagonalisation.  Note that $\hat U$ is no longer compatible and we have to 
use spatial inversion symmetry to demonstrate that the Mazur contribution 
from the Hamiltonian now vanishes.  We employ this field in our calculations 
as a natural parameter to contrast the different behaviours and assess whether 
or not half-filling is a pathological case.

We are now led to a crucial issue, that of the {\it compatibility} of the 
current with the conservation laws:  We have chosen to complete the 
integrability of our model using the $z$-component of total-spin (which would 
be forced in the presence of the field).  If, instead, we chose to complete 
the {\it isotropic} case, $\Delta $=1, with the $x$-component of the 
total-spin (or used $\hat U$ itself for the general case) then $\hat U$ remains 
a symmetry and the current operator is necessarily off-diagonal.  For this case 
there is no Mazur contribution and the current operator is {\it incompatible} 
with the maximal set of mutually commuting conservation laws.  Recognising that 
this non-Abelian aspect to the conservation laws can be critical, we now return 
to the case where we employ the $z$-component of total-spin and the current is 
compatible with the conservation laws.

The monodromy matrix provides us with a conservation law for each value of 
$\lambda $, and by varying $\lambda $ we have access to a set of independent 
conservation laws.  Indeed, for a chain of length $N$ we have $N$ linearly 
independent conservation laws.  Unfortunately, these are {\it not} the pure 
local conservation laws but for each order an additional polynomial of lower 
order local conservation laws is incorporated into these conservation laws.  
In the issue of saturation this does not cause problems, but making particular 
deductions about the local conservation laws is usually made impractical.  
Although we generate a `complete' set of conservation laws there is an 
important subtlety.  A particular conservation law can be used to generate 
more:  Any polynomial combination of a conservation law is another such law 
and as we have seen, the projection operators, which project onto eigenspaces, 
are a complete set of conservation laws which can be generated by any 
particular conservation law.  Further, it should be appreciated that for $N$ 
sites we have a state-space which scales as $2^N$ and is {\it exponential} in 
system size.  Usually, the number of derived conservation laws is exponential 
in the system size and the total number of conservation laws required to get a 
complete set is disturbingly large even for quite a small system.

We perform a sequence of numerical calculations:  For a chain of length $N$ 
we choose $N$ distinct values of $\lambda $ and create $N$ distinct 
monodromy matrices.  We then calculate our four quantities as a function of 
applied field for the set of all $N$ such conservation laws.  Note that we 
choose to {\it orthogonalise} these conservation laws because of the 
additional stability this offers for the case when the conservation laws are 
{\it over-complete.}  We then create new conservation laws by taking the 
$N(N+1)/2$ distinct products of these monodromy matrices and include these 
in a larger calculation with both linear and quadratic combinations and 
calculate a second estimate to the Mazur inequality. 
We then calculate yet further 
conservation laws by taking the $N(N+1)(N+2)/6$ distinct triple products of 
these monodromy matrices and including them into a third Mazur estimate. 
We took this procedure to quartic order but proceeded no further.  The 
results for a couple of representative cases, one metallic and one insulating, 
are provided in figure 1.
\begin{figure}
\includegraphics[height=7.2 cm, width=8.4 cm]{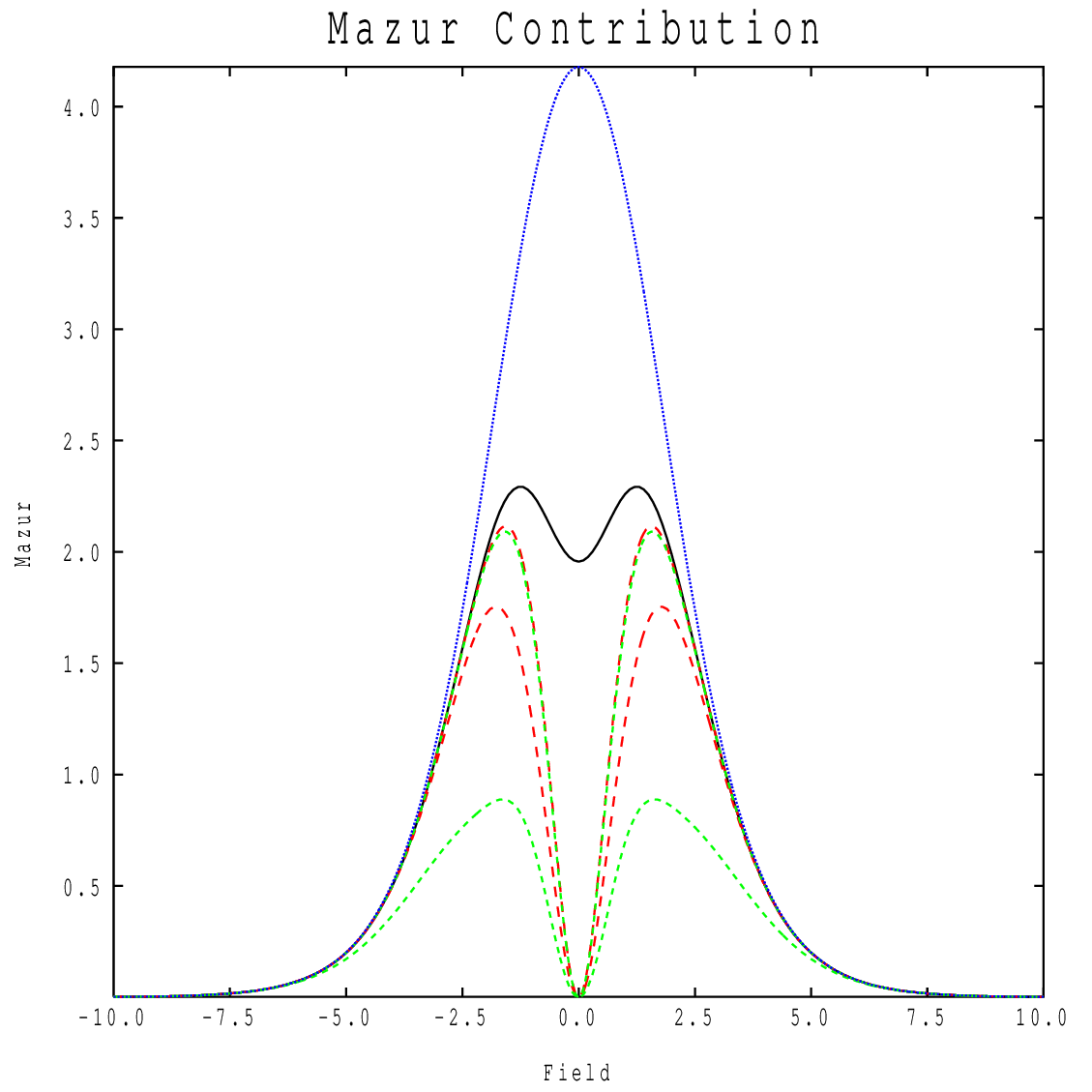}
\includegraphics[height=7.2 cm, width=8.4 cm]{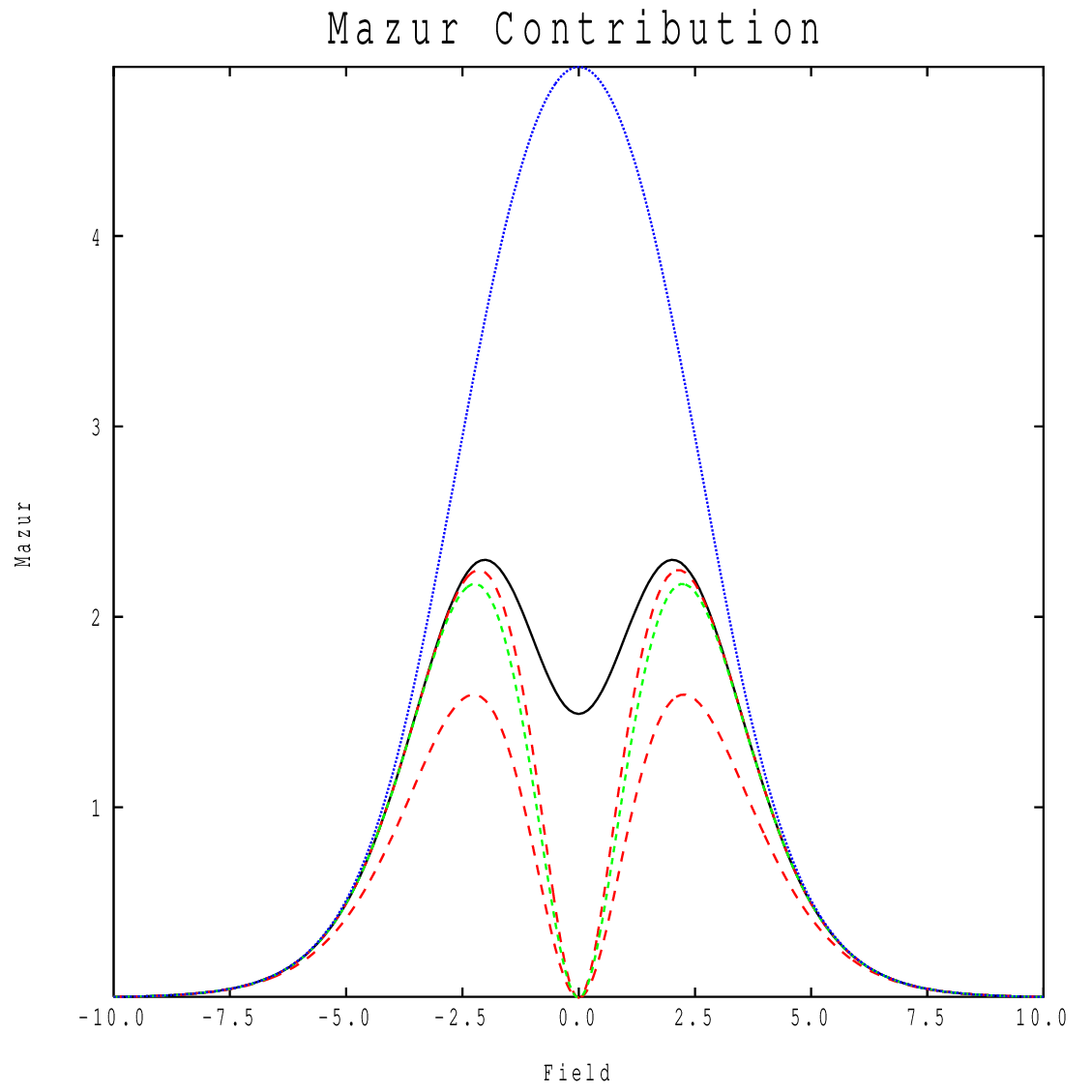}
\caption{\label{fig:1} Long-time residuals and estimates to the Mazur 
inequality.  
We offer the original current, the exact long-time residual and a sequence 
of three estimates coming from the conservation laws.  The third 
calculation exhausts all the available conservation laws.  $N$=8, $\delta $=1, 
$\beta $=1 and (a) $\Delta <$1 (b) $\Delta >$1.}
\end{figure}
The conundrum is clearly observable:  The contribution to the Mazur inequality 
from the conservation laws smoothly vanishes at zero field and some of the 
long-time residual is lost.  For a finite system we can {\it exhaust} the 
attainable conservation laws and this observed loss is genuine!  Fortunately, 
the resolution of this issue is straightforward:  One important conservation 
law was missing from our analysis, $\hat S^z$, the $z$-component of 
total-spin.  The monodromy matrix is only non-degenerate when we {\it 
additionally} respect this symmetry too.  Including this conservation law 
into our previous calculations, by including $\hat S^z$ times the previous 
conservation laws as `independent' laws, provides figure 2.
\begin{figure}
\includegraphics[height=7.2 cm, width=8.4 cm]{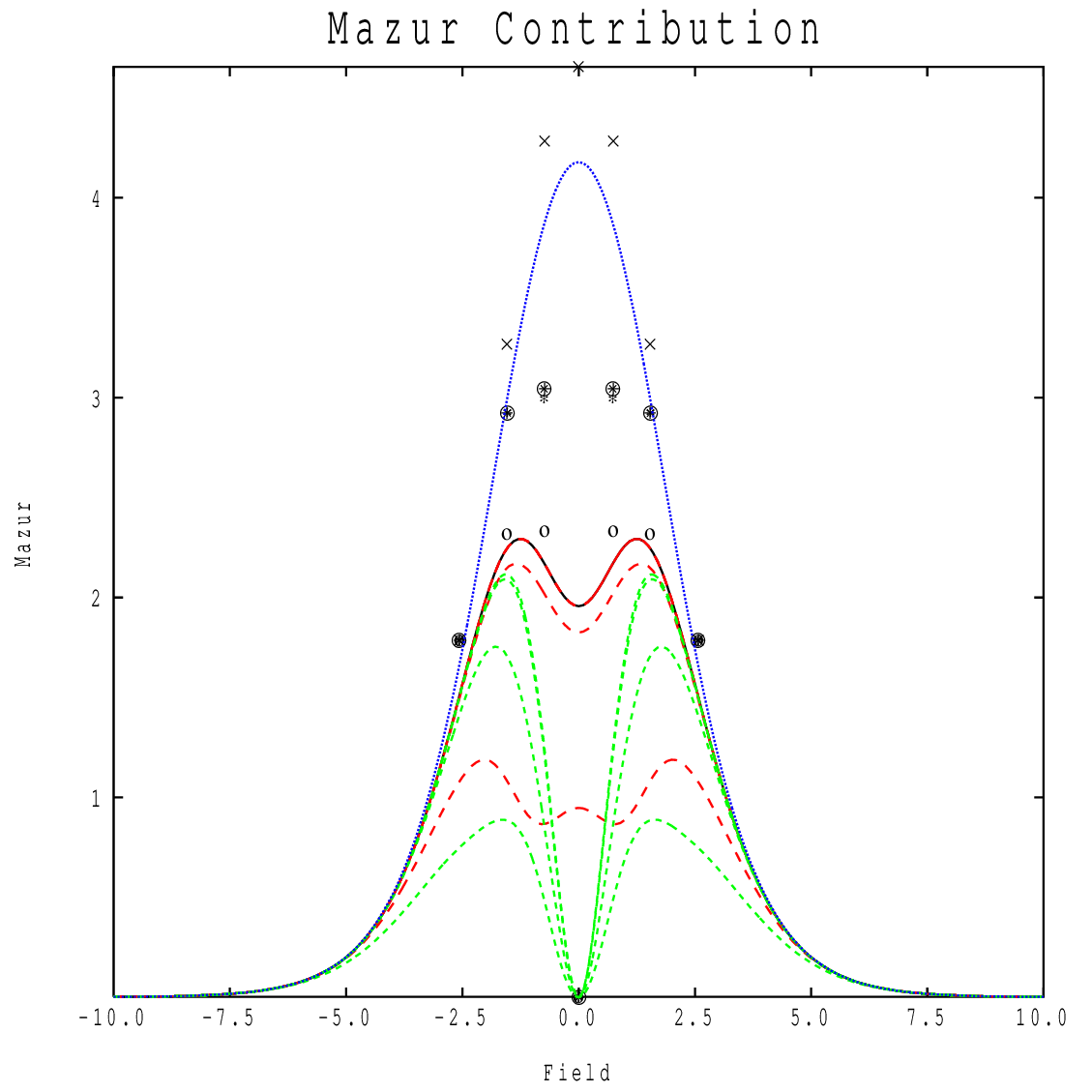}
\includegraphics[height=7.2 cm, width=8.4 cm]{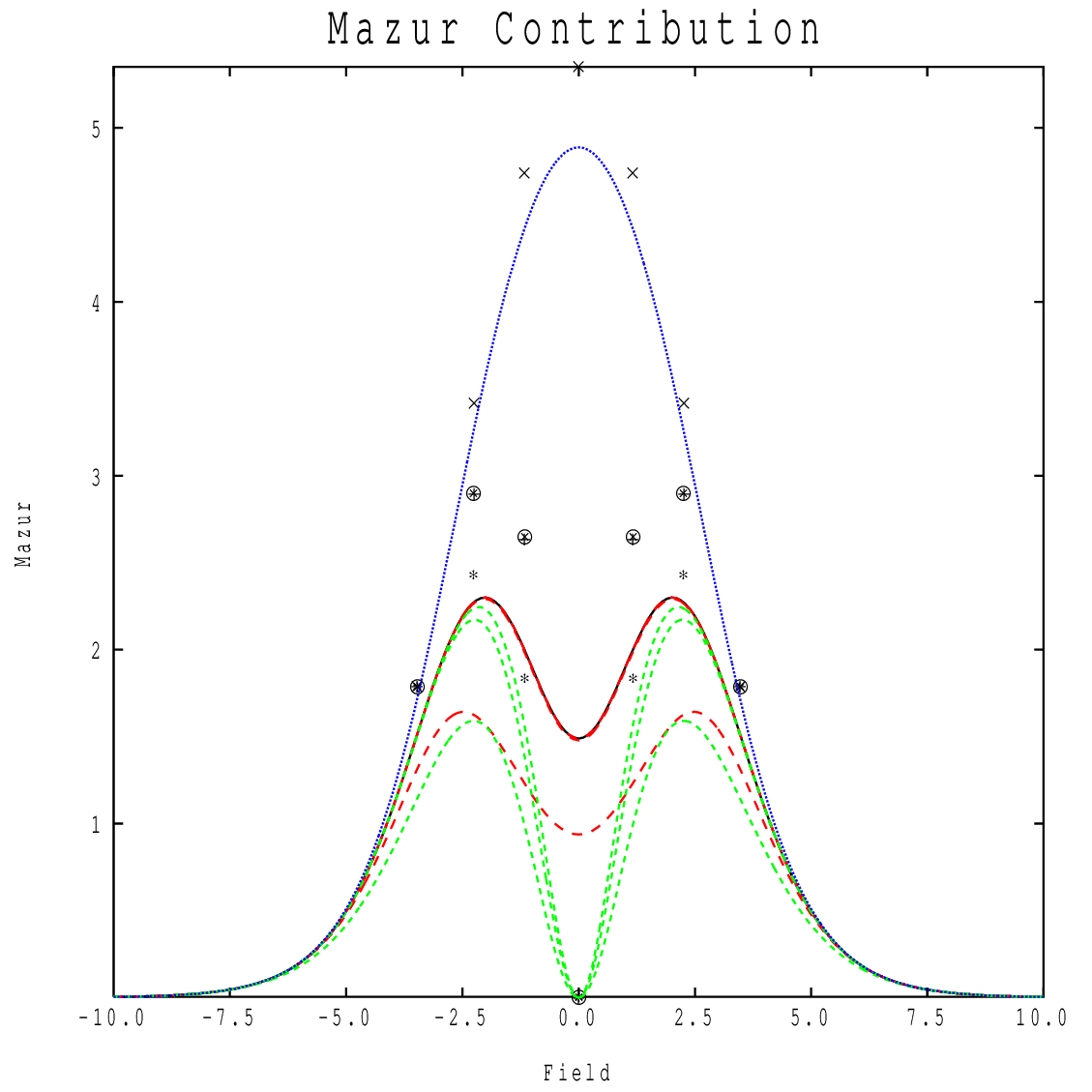}
\caption{\label{fig:2} Long-time residuals and Mazur estimates.  
We offer the original current, the exact long-time residual and a sequence 
of four estimates coming from the conservation laws, both with and 
without the total-spin $\hat S^z$ included.  The fourth 
calculation exhausts all the available conservation laws.  $N$=8, $\delta $=1, 
$\beta $=1 and (a) $\Delta <$1 (b) $\Delta >$1.}
\end{figure}
The problem at `half-filling' has been alleviated and for 
small systems {\it all} the conservation laws can be exhausted and we can 
achieve the full long-time residual.

For the particular case of the XXZ model and the standard spin current, we now 
have an answer to our initial question:  an initial current does decay as a 
function of time, except at $\Delta $=0 or $\infty $, and the long-time 
residual is {\it exhausted} by the known conservation laws.  In order to 
exhaust this residual one {\it must} include $\hat S^z$ into the list of 
conservation laws, because without this law the residual is {\it not} 
exhausted and indeed at half-filling the local conservation laws offer {\it 
nothing.}  Our results are for {\it finite} systems and there are {\it 
definite} problems in the thermodynamic limit. 
It is `known'\cite{z,kluemper}, from the 
Bethe Ansatz, that at half-filling the long-time residual shows interesting 
behaviour:  For $\Delta <$1 we have a finite residual that smoothly vanishes 
as $\Delta \mapsto $1 and vanishes for $\Delta >$1.  A {\it single} local 
conservation law must find difficulty in providing this behaviour and it is 
our need for the combination of $\hat S^z$ with the local conservation laws 
which overcomes this difficulty.

Since any particular value of $\lambda $ generically provides a non-degenerate 
monodromy matrix, we could have used a {\it single} such conservation law to 
exhaust the conserved current.  We have demonstrated that this is so on very 
small systems, but the similarity of the laws provided by powers of this one 
law lead to unpleasant instability issues numerically, which, together with 
the added control of completeness issues, made our chosen procedure more 
effective.

Since we are dealing with the physical current in the system, but the 
fundamental theory should be applicable to an arbitrary operator, one can ask 
if the results are in any way special for the physical current.  We repeated 
our calculations for the longer-range `currents'
\begin{equation}
\hat j_n\equiv (-i)\sum _j\left( \hat S^+_j\hat S^-_{j+1+n}-\hat S^-_j\hat 
S^+_{j+1+n}\right) ,
\end{equation}
where $\hat j_0$ is the physical current and $\hat j_1$ and $\hat j_2$ are 
accessible to our available system sizes and they exhibit essentially no new 
physics at all:  The operator $\hat U$ is still available to show that the 
local conservation laws offer nothing in the absence of a field and the 
analogue plots to figure 2 are equivalent in all respects.

What can we say in general?  At the formal level our results are clear:  We see 
four natural quantities, $I_0\ge I_1\ge I_2\ge I_3$, and in principle they are 
all distinct.  The initial current, controlled by $I_0$, only remains 
undiminished if the current operator is a conservation law itself, as happens 
for the case $\Delta $=0.  The long-time residual, controlled by $I_1$, may be 
considered to come from conservation laws, but only if we allow the inclusion 
of non-Abelian laws.  If we have a set of mutually commuting conservation laws, 
then we only exhaust the long-time residual when the previously considered 
conserved current is {\it the identity} in the basis which diagonalises the 
conservation laws and also conserves any residual degeneracy.

We have shown that the symmetry at zero field causes the local conservation 
laws to become independent from the conserved current and we have resolved this 
by including the $z$-component of total-spin to successfully represent the 
entire conserved current, but all this was for finite systems:  What happens 
in the thermodynamic limit?  To exhaust the current conservation law we need 
to generate the long-time residual to machine accuracy from a Mazur calculation 
and this we can achieve with all accessible systems which do not have to be 
large to be convincing.  To analyse the thermodynamic limit we need to 
finite-size scale and our relatively small systems are only indicative and 
not at all convincing.  Nevertheless, we believe that in the thermodynamic 
limit the point at zero field becomes {\it singular} and that at all other 
fields the local conservation laws {\it exhaust} the conserved current.  One 
analytical consideration is that in the thermodynamic limit we should get the 
same answer if we canonically fix the magnetisation or grand-canonically fix 
the field:  Since when the magnetisation is fixed there is no difference 
between the conservation laws with and without magnetisation (except perhaps 
when the magnetisation vanishes), then there can only be one quantity 
generated by these laws, the full conserved current.  We have incorporated 
the canonical calculations into figure 2 and it is plausible that they 
converge to the long-time residual.  We have also finite-size scaled the 
fraction of the long-time residual provided by the conservation laws in 
figure 3
\begin{figure}
\includegraphics[height=7.2 cm, width=8.4 cm]{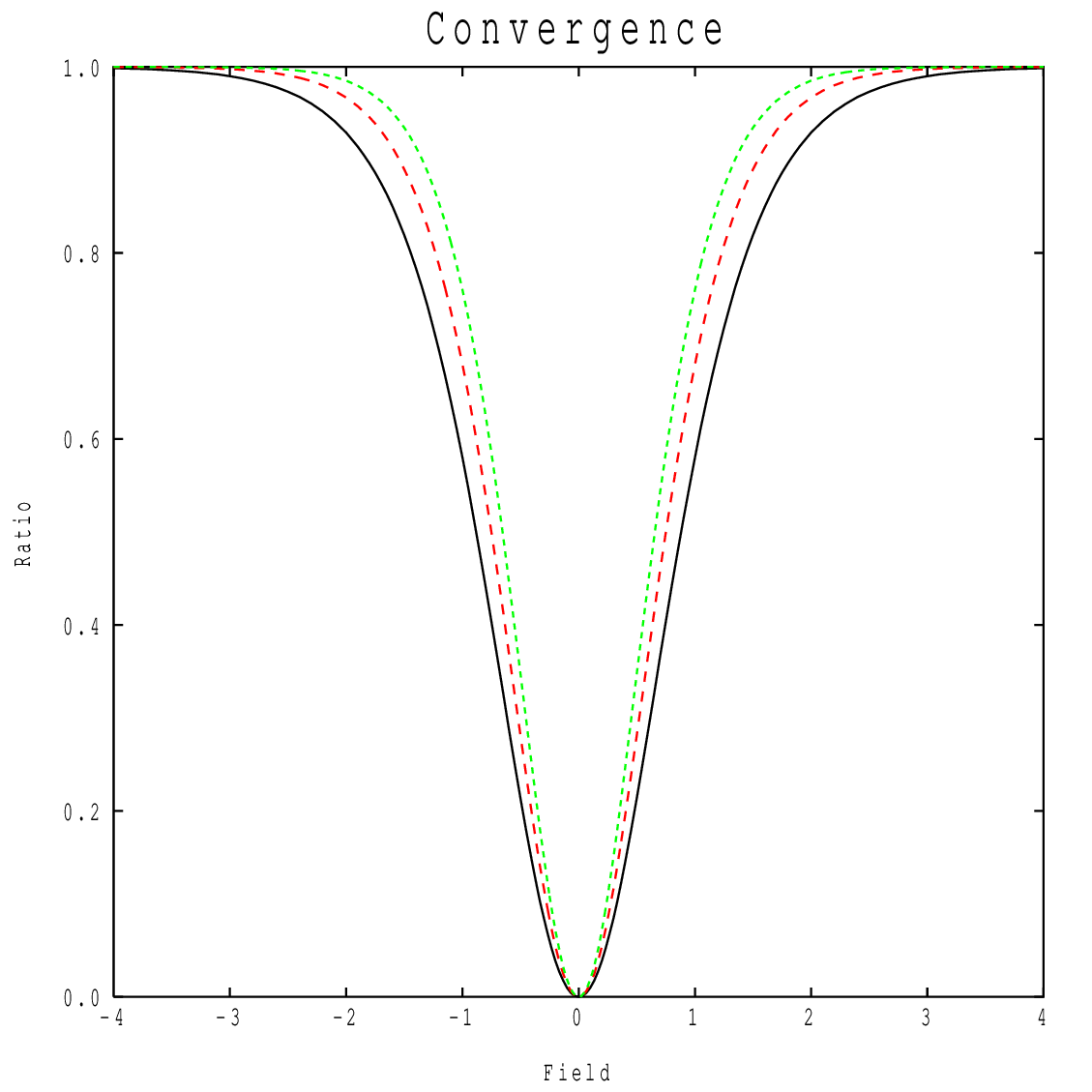}
\caption{\label{fig:3} Overlay of the ratios of exhausted Mazur estimates 
without the total-spin $\hat S^z$ included.  $N$=4,6,8, $\delta $=1, 
$\beta $=1 and $\Delta <$1.}
\end{figure}
and it is easiest to believe that in the thermodynamic limit the point at 
zero magnetisation is an isolated puncture.

Although we appear to have a large number of degrees of freedom in our 
calculations (4096 for $N$=12 for example) a brief analysis of the 
available degeneracy might suggest that our systems are {\it too small} 
to be relevant:  Once we have extracted the $S^z$ degeneracy employing a 
field, although at first sight there appears to be a huge residual degeneracy, 
in fact this degeneracy is almost all lifted by inversion symmetry and the 
further residual degeneracy is woefully small!  Indeed, we may employ the 
first non-trivial local conservation law, the energy current\cite{znp}, to 
lift the inversion symmetry and up to $N$=11 this {\it fully} diagonalises 
the system.  Additional degeneracy emerges for $N$=12, but this is much 
less than is naively expected:  We might expect that at order $N$ the $N$ 
distinct conservation laws were {\it functionally independent,} but for 
systems up to $N$=11 all $N$ laws are describable as functions of the 
Hamiltonian, the energy current and the $z$-component of total-spin!  The 
role of degeneracy in the thermodynamic limit is currently not accessible 
to computers, even at the level of a hint!

For the Heisenberg model, we have established that for a finite system we can 
expect that
\begin{equation}
\langle \hat j(\infty )\hat j(0)\rangle =\langle \hat j_c\hat j_c\rangle ,
\end{equation}
with
\begin{equation}
\hat j_c={\cal F}[\hat S^z,\hat C^a]
\end{equation}
for some function of the relevant conserved quantities.  The final issue is 
whether we can use this knowledge to {\it prove} the existence of a non-trivial 
long-time residual in the thermodynamic limit.  Employing the first non-trivial 
conservation law, the heat current, it was shown that there is a non-vanishing 
residual at finite temperature for all fillings {\it except} 
half-filling\cite{znp}.  
As we have seen, at half-filling we also require to use the conservation law 
$\hat S^z$, so it seems plausible that we might be able to use a combination 
of both the heat current and the total spin to find the residual at 
half-filling.  Unfortunately, this does not appear to be true and a 
verification of the long-time residual at half-filling using 
the Mazur inequality must await further investigation!

A natural technique is to Taylor expand the general conservation law and create 
a trivial conservation law from which the long-time residual can be deduced 
from the Mazur inequality
\begin{equation}
\langle \hat j(\infty )\hat j(0)\rangle \ge \frac{\langle \hat C^\dagger \hat j
(0)\rangle \langle \hat j(0)\hat C\rangle }{\langle \hat C^\dagger 
\hat C\rangle }.
\end{equation}
Clearly all the $\hat C^a$ have the wrong symmetry and $\hat S^z$ itself 
also has the wrong symmetry and so we might like to start from the law
\begin{equation}
\hat C\equiv \hat S^z\hat C_1 ,
\end{equation}
for example.  We can analyse this proposal in detail for the limit $\Delta $=0 
where the Jordan-Wigner transformation reveals the solution as a 
non-interacting free-electron gas.  Employing
\begin{eqnarray}
\hat j\equiv 2\sum _k\sin kf^\dagger _kf_k , &&\nonumber \\
\hat C^1\equiv \sum _k\sin 2kf^\dagger _kf_k , &&\\
\hat S^z\equiv \sum _k\left[ f^\dagger _kf_k-\frac{1}{2}\right] , &&\nonumber 
\end{eqnarray}
we can verify that
\begin{equation}
\langle \hat j\hat C^1\hat S^z\rangle \equiv 2\sum _k\sin k\sin 2k\; n_k
[1-n_k][1-2n_k] ,
\end{equation}
with
\begin{equation}
n_k=\langle c_k^\dagger c_k\rangle =\frac{1}{1+e^{\beta \epsilon _k}}
\end{equation}
at half-filling and does {\it not} vanish, as expected.  Unfortunately, in the 
thermodynamic limit this conservation law offers nothing because it scales 
with the length of the chain but
\begin{equation}
\langle (\hat C^1)^2(\hat S^z)^2\rangle \mapsto \langle (\hat C^1)^2
\rangle \langle (\hat S^z)^2\rangle 
\end{equation}
scales with the {\it square} of the chain length rendering the contribution 
negligible!  Note that the case $\Delta $=0 is special:  There is a {\it 
second} ladder of conservation laws\cite{gm} which includes the current 
operator itself.  When $\Delta \neq $0 the conservation laws
\begin{eqnarray}
C_{2m}\equiv \sum _k\cos (2m+1)kf^\dagger _kf_k , &&\nonumber \\
C_{2m-1}\equiv \sum _k\sin 2mkf^\dagger _kf_k&&
\end{eqnarray}
smoothly connect to the local conservation laws whilst the laws
\begin{eqnarray}
D_{2m}\equiv \sum _k\cos 2mkf^\dagger _kf_k , &&\nonumber \\
D_{2m-1}\equiv \sum _k\sin (2m-1)kf^\dagger _kf_k&&
\end{eqnarray}
are broken (except $D_0$ which is not part of the ladder).  Noting the 
symmetries of these classes, the law
\begin{equation}
\hat C\equiv \hat S^z\left[ c^\dagger _qc_q-c^\dagger _{\pi -q}c_{\pi -q}
\right] 
\end{equation}
contributes from the appropriate sector and yields a {\it finite} Mazur 
contribution.  
This law arises from a linear superposition of the local 
conservation laws but it itself is clearly long-range.

One of the central current issues is that of finding a conservation law for 
which a single application of the Mazur inequality would provide 
incontrovertible proof of the existence of a long-time residual.  We already 
know that this must be non-trivial for the current model, since the 
long-time residual is known to vanish for $\Delta >$1 and hence any 
prospective conservation law must respect this fact, a very non-analytic 
prospect.  If we consider the generation of part of the long-time residual 
using a single conservation law, then we encounter an immediate problem if we 
attempt to employ {\it local} laws.  If we consider the conservation law
\begin{equation}
\hat S^z\hat C ,
\end{equation}
where we {\it require} to use a local conservation law {\it and} the 
$z$-component of total-spin, then we may replace this with
\begin{equation}
\left[ \hat S^z-\langle \hat S^z\rangle \right] \left[ \hat C-\langle \hat C
\rangle \right] -\langle \left[ \hat S^z-\langle \hat S^z\rangle \right] \left[ 
\hat C-\langle \hat C \rangle \right] \rangle ,
\end{equation}
since neither operator has an overlap with the current on its own.  Now 
the Mazur inequality tells us that
\begin{widetext}
\begin{equation}
\frac{\langle \hat j(\infty )\hat j(0)\rangle }{\langle \hat j(0)\hat j(0)
\rangle }\ge \frac{\langle \left( \hat j-\langle \hat j\rangle \right) \left( 
\hat S^z-\langle \hat S^z\rangle \right) \left( \hat C-\langle \hat C\rangle 
\right) \rangle \langle \left( \hat C-\langle \hat C\rangle \right) \left( 
\hat S^z-\langle \hat S^z\rangle \right) \left( \hat j-\langle \hat j\rangle 
\right) \rangle }{\langle \left( \hat j-\langle \hat j\rangle \right) 
\left( \hat j-\langle \hat j\rangle \right) \rangle \langle \left( 
\hat S^z-\langle \hat S^z\rangle \right) \left( \hat S^z-\langle \hat 
S^z\rangle \right)  \left( \hat C-\langle \hat C\rangle \right) 
 \left( \hat C-\langle \hat C\rangle \right) \rangle } ,
\end{equation}
\end{widetext}
and note that, employing notation
\begin{equation}
\delta \hat A\equiv \hat A-\langle \hat A\rangle ,
\end{equation}
we have
\begin{eqnarray}
&&\langle \delta \hat S^z\delta \hat S^z\delta \hat C\delta \hat C\rangle =
\langle \delta \hat S^z\delta \hat S^z\rangle \langle \delta \hat C
\delta \hat C\rangle \nonumber \\&&+\langle \left[ \delta \hat S^z\delta \hat 
S^z-\langle \delta \hat S^z\delta \hat S^z\rangle \right] \left[ \delta \hat 
C\delta \hat C-\langle \delta \hat C\delta \hat C\rangle \right] \rangle ,
\end{eqnarray}
where the first term is likely to dominate.  These ideas are essentially true 
whatever the operator $\hat C$, but if $\hat C$ is {\it local} and the 
correlations in the system are {\it short-range} then we can deduce 
something from this:  If we have that
\begin{equation}
\hat A\equiv \sum _i\hat A_i
\end{equation}
for all such local operators then
\begin{equation}
\langle \delta \hat A\delta \hat B\rangle =\sum _{ij}\langle \delta \hat A_i
\delta \hat B_j\rangle ,
\end{equation}
and if $i$ and $j$ become far apart we might expect the correlations to 
become irrelevant.  This correlation function might then be expected to scale 
{\it linearly} with system size.  Similarly, including a $\hat C$ then
\begin{equation}
\langle \delta \hat A\delta \hat B\delta \hat C\rangle =\sum _{ijk}\langle 
\delta \hat A_i\delta \hat B_j\delta \hat C_k\rangle 
\end{equation}
also might be expected to scale {\it linearly} with system size.  The 
consequence of this is that {\it any local} contribution to the Mazur 
inequality would be 
expected to {\it vanish} in the thermodynamic limit.  In practice, the 
correlations are likely to be {\it power-law} in nature, but the physical 
idea behind the failure remains.

The hunt for an elementary conservation law that will provide a non-vanishing 
Mazur contribution at zero field probably necessitates non-local conservation 
laws, but we should remember that the monodromy matrix expanded in the manner 
that we employ {\it does} yield non-local laws\cite{nonlocal} and further we 
are {\it complete} for our finite systems so we employ an {\it arbitrary} 
linear superposition.  Based on the previous limit of $\Delta \mapsto $0 we 
might suspect that it is only an arbitrary superposition over all lengths that 
is crucial and {\it not} the non-linear combinations of laws.  Our final 
calculation offers a test of this idea through a finite-size scaling 
calculation of the fraction of the long-time residual provided by the linear 
conservation laws, the quadratic conservation laws and so on, order by order, 
in figure 4,
\begin{figure}
\includegraphics[height=7.2 cm, width=8.4 cm]{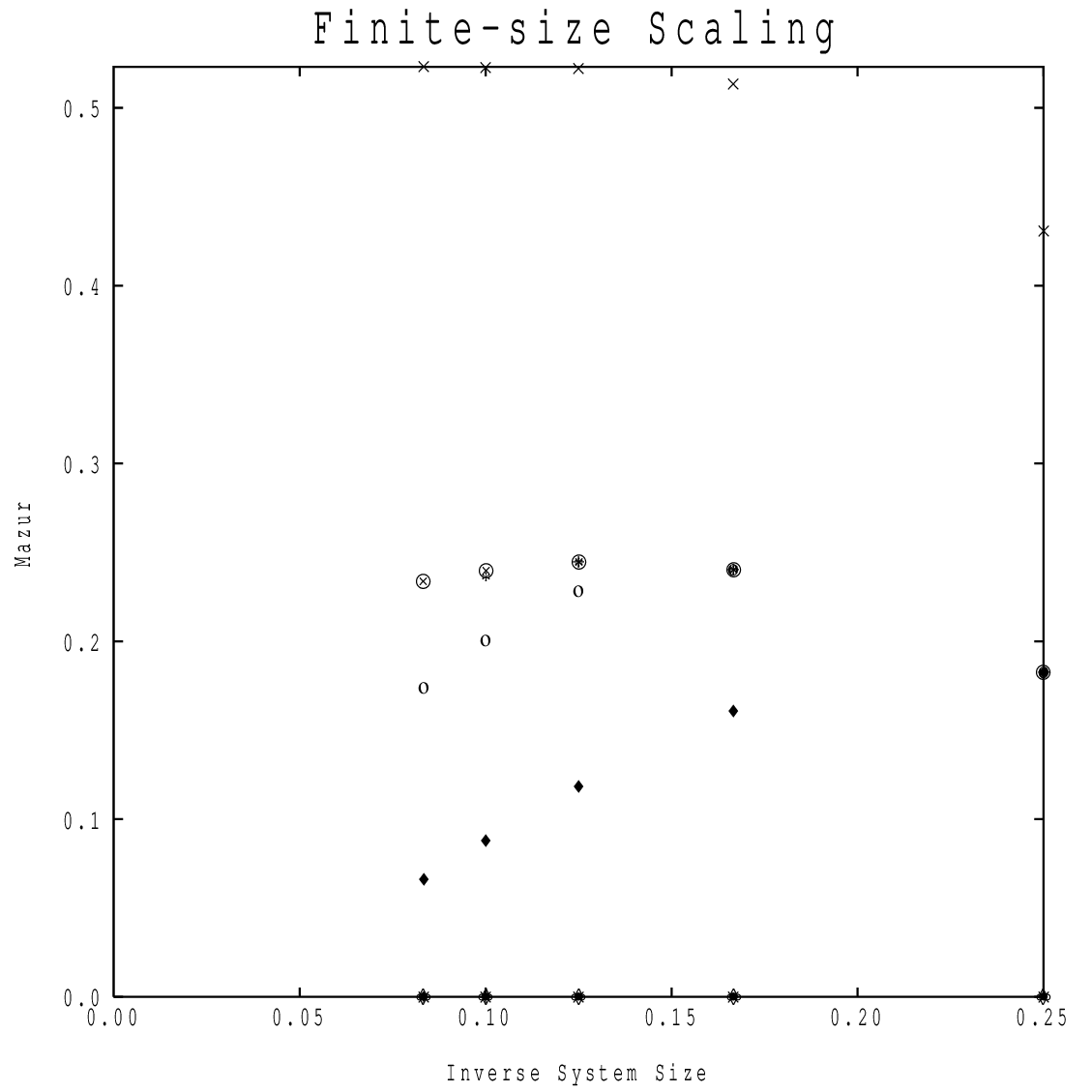}
\caption{\label{fig:4} Finite-size scaling plot of the fraction of the 
long-time residual achieved by the linear contribution to the conservation 
laws for the case $N$=4,6,8,10, $\delta $=1, $\beta $=1 and $\Delta <$1.}
\end{figure}
and unfortunately the contribution scales away, reinforcing the fear that a 
truly non-analytic function of the conservation laws is required.

\section{Conclusions}

Integrable systems can have anomalous conductivity, with currents flowing 
indefinitely once started.  As suggested by Suzuki\cite{suzuki}, 
this long-time residual 
can be thought of as some part of the current operator actually {\it being} 
one of the conservation laws of the system.  We agree with this statement but 
refine one of the issues arising:  Given a set of conservation laws, when is 
the conserved current either partially or wholly described by them.  One 
might anticipate that a `complete set of mutually commuting' conservation 
laws would generate {\it all} possible conservation laws, but this is 
incorrect:  In general, additional independent conservation laws can exist 
with non-trivial commutation with the complete set.  This is {\it not} just 
a formal issue, as can be seen by example:  The isotropic Heisenberg model 
with the standard spin current.  The local conservation laws do {\it not} 
completely diagonalise the system and there are a variety of ways of 
completing the conservation laws.  We can simply diagonalise one of the 
components of total-spin as well, for example.  If we choose to use the 
$z$-component of total-spin then we can generate the full conserved current 
but if we choose either the $x$- or $y$-component of total spin then we can 
generate no long-time residual whatsoever!  The current must be {\it 
compatible} with the complete set of mutually commuting conservation laws.

Using numerics on finite systems, we show how to generate the current 
conservation law for the XXZ model:  If we combine the local conservation 
laws\cite{gm} 
with the $z$-component of total spin then we can exhaust the long-time 
residual and consequently form the complete conserved current.  Since 
the local conservation laws yield no long-time residual at all, in the 
absence of a field, any conservation law which has an overlap with the 
current must involve both the local conservation laws {\it and} the 
$z$-component of total-spin.

The final issue addressed is as to the reason for the failure of any 
elementary conservation laws to provide even {\it part} of the conserved 
current for the case with no field.  We observe that since we require to 
employ both a local conservation law and the $z$-component of total-spin then 
for short-range interactions we need a coincidence of {\it three} local 
operators to yield a contribution and a consequent loss of this contribution in 
the thermodynamic limit.  It is not easy to use the conservation laws to 
demonstrate a long-time residual in practice!

\begin{acknowledgments}
We wish to acknowledge useful discussions with J.M.F. Gunn and C.A. Hooley. 
X.Z would like to acknowledge financial support 
by the E.U. grant MIRG-CT-2004-510543.
\end{acknowledgments}


\end{document}